\documentclass[12pt,twoside]{article}
\usepackage{cmp2e}
\usepackage{graphicx}
\title{Ising antiferromagnet with mobile, pinned and quenched defects}
\author{W.~Selke, M.~Holtschneider, and R.~Leidl}
\address{Institut f{\"u}r Theoretische Physik, Technische Hochschule, 52056 Aachen, Germany}

\begin{document}

\maketitle

\begin{abstract}
  Motivated by recent experiments on
  (Sr,Ca,La)$_{14}$Cu$_{24}$O$_{41}$, a two--dimensional Ising 
  antiferromagnet with mobile, locally pinned
  and quenched defects is introduced and analysed using
  mainly Monte Carlo techniques. The interplay between the 
  arrangement of the defects and the magnetic ordering
  as well as the effect of an external field are
  studied.

\pacs 05.10.Ln, 05.50+q, 74.72.Dn, 75.10.Hk
\end{abstract}

\section{Introduction}

Several interesting low--dimensional
magnetic properties arise from the CuO$_2$ chains in
(Sr,La,Ca)$_{14}$Cu$_{24}$O$_{41}$. Pertinent
experimental findings \cite{miz,amm,mat,klin}
motivated recent theoretical studies on
two--dimensional Ising antiferromagnets with
defects \cite{selke,holt1,hls,kroll}.

In particular, a simple Ising model on a square
lattice with mobile defects has been introduced \cite{selke}, with
the chain direction corresponding to one of the
axes of the lattice. The
spins, $S_{(i,j)} = \pm 1$ at lattice site $(i,j)$, correspond
to the magnetic Cu$^{2+}$ ions, and the defects, $S_{(i,j)}= 0$, to
those Cu ions which are believed to be
spinless due to holes (Zhang--Rice singlets). In the model neighbouring
spins are supposed to be 
coupled ferromagnetically in each chain and
antiferromagnetically in
adjacent chains. In addition, each
pair of next--nearest neighbour spins in the same chain
separated by a defect is
presumed to interact strongly
antiferromagnetically. The (mobile) defects are allowed to
hop to neighbouring sites in the chains, with the 
energy barriers of these moves given merely by the magnetic 
couplings.

Without defects, the model describes a
two--dimensional Ising antiferro- or metamagnet, with ferromagnetic
ordering in the chains and antiferromagnetic ordering
between the chains in the low--temperature phase \cite{rott}. The defects 
form, at low temperatures, nearly straight
stripes, perpendicular to the CuO$_2$ chains, separating
antiferromagnetic domains. The coherency of the stripes gets lost
at a phase transition of first order \cite{selke,holt1}. 

To mimic the possible pinning of the defects, due to the
La ions, variants of the model may be considered with
local pinning positions at periodic \cite{holt1,hls} or random sites. For
indefinitely large pinning strength, the defects
will be quenched at fixed sites. Obviously, pinning or 
even quenching may affect the stability of the defect
stripes, the magnetic ordering, and related phase
transitions.

Note that the model, albeit being motivated experimentally, is
thought to be of genuine theoretical
interest as well, describing the interplay of magnetic
structures and defect arrangements. In that
respect, the analysis of the model belongs to the intriguing
studies dealing with various aspects on randomness in
magnets, ranging, say, from site diluted ferromagnets to
spin glasses \cite{binder,sts,folk,janke,cala,vojta}.

Concerning the interpretation of experiments on
(Sr,La,Ca)$_{14}$Cu$_{24}$O$_{41}$, attention may be drawn also to 
current theoretical analyses based on two--dimensional anisotropic
Heisenberg models \cite{mat,leidl1,leidl2}.

\begin{figure}
\begin{center}
\includegraphics[width=0.55\linewidth]{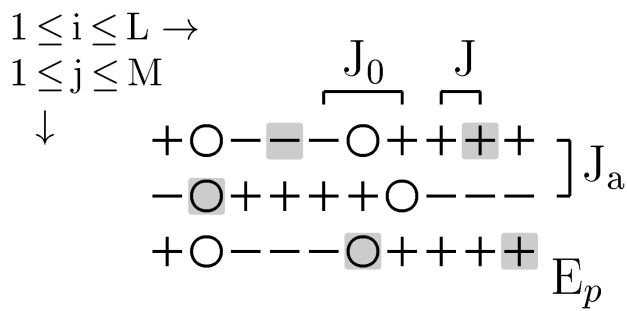}
\end{center}
\caption{Sketch of the model. Pinning sites are indicated by
shadowed squares.}
\end{figure}

The article is organized as follows: In the next section, the
Ising model with defects is introduced. Results are discussed
in Section 3, reviewing as well as illustrating
previous, recent findings and presenting new results on the
full model with quenched defects. A brief summary concludes
the paper.

\section{Model}

The Ising model with mobile, pinned or quenched defects is defined on
a square lattice
with one axis corresponding to the chain direction, say,
the horizontal, $i$--axis, as shown in Fig.~1. Each
lattice site, $(i,j)$, is occupied either by a spin, $S_{(i,j)}= \pm 1$, or
by a defect, $S_{(i,j)}= 0$. The concentration of
defects is fixed to be ten percent of the lattice sites, as it seems to
be the case in La$_5$Ca$_9$Cu$_{24}$O$_{41}$. For simplicity, we
shall assume the same concentration of defects in each chain. The
interactions, as sketched in Fig.~1, are as follows. Neighbouring spins
are coupled ferromagnetically, $J >0$, along the chains, and
antiferromagnetically, $J_a < 0$, perpendicular to
them. In addition, next--nearest neighbouring spins
in the same chain separated by a defect interact
antiferromagnetically, $J_0 <0$, as suggested by experiments. A local
pinning potential, $E_p(i_p,j_p)$ at fixed sites $(i_p,j_p)$ may act
upon defects, reducing the energy by the pinning strength. Here
we shall assume that the concentration of pinning sites is identical
to the concentration of defects, with the same number of pinning
sites and defects in each chain. Furthermore, the pinning
strength $E_p$ is taken to be the same at
each pinning site. If $E_p= 0$, the
defects are called mobile, at finite values of $E_p > 0$, they are
pinned, and at $E_p \longrightarrow \infty$, the defects
are quenched. For finite pinning strength, defects are allowed
to diffuse along the chains, keeping a minimal distance of two lattice
spacings.

When $J$ and $|J_0|$ are large compared to $|J_a|$, as suggested
by experiments, one may arrive at
the 'minimal version' of the model, where spins along a
chain are assumed to have the same sign between two defects, reversing sign
at a defect. The only relevant energy parameter
is then $J_a$ \cite{selke}.

In the following, we shall present properties of the 
minimal model with mobile defects and defects pinned
at periodic pinning lines perpendicular to the chains, extending
recent work \cite{selke,holt1,hls,kroll,holt2}. 

The full model, where all couplings are finite, will be
discussed without pinning as well as with defects quenched
at periodic pinning lines and
at random sites. In the following, we take $J_a= -0.3J$ and
$J_0= -6.25J$, choices based on experimental
input \cite{selke}. 

Analyses are based on ground state considerations, the
free--fermion method to describe thermal properties
at low temperatures, the transfer matrix approach, and
Monte Carlo simulations.

Simulations are rather demanding, because
of, usually, slow fluctuations due to the
defects. Typically, runs over at least $10^6$ Monte Carlo
steps per site were performed, averaging then over several
realizations, simulating lattices with linear dimension $L$
in the $i$ direction, and $M$ in the $j$ direction, see
Fig.~1. In the following we take $L=M$. In
the case of pinning at random sites, an ensemble average has
to be taken. Finally, different system sizes have to be
simulated especially to determine accurately the phase
transition temperatures.

Physical quantities of interest \cite{selke,holt1} include the
specific heat, $C$, the
susceptibility, $\chi$, and
spin correlation functions parallel to the chains,
\begin{eqnarray}
 G_1(i,r)= \left(\sum\limits_{j} \langle S_{i,j} S_{i+r,j} \rangle \right)/M,
\end{eqnarray}
and perpendicular to the chains,
\begin{eqnarray}
 G_2(i,r)= \left(\sum\limits_{j} \langle S_{i,j} S_{i,j+r} \rangle \right)/M,
\end{eqnarray}

Without pinning as well as in the case of defects quenched at random sites,
the defect positions are expected to be uncorrelated, so that there is
translational invariance with the spin correlations not
depending on $i$. Note  that in the thermodynamic
limit ($L,M \longrightarrow \infty$) for
infinitely large distance $r$ the correlations determine the
(sublattice) magnetization. 

We also calculated less common microscopic quantities which describe
the stability of the defect stripes and the ordering of the defects in the
chains \cite{selke,holt1}, including the average
minimal distance $d_m$ between each
defect in chain $j$, at position $(i_d,j)$, and those in the next
chain, at $(i_d',j+1)$, and the cluster distribution $n_d(l)$ denoting
the probability of a cluster with $l$ consecutive spins of equal sign
in a chain (as considered, e.g., in percolation
theory \cite{stauffer}). Our main
emphasis will be on pairs of defects with $l= 1$. Finally, it
turned out to be quite useful to visualize the microscopic spin
and defect configurations as encountered during the simulation.

\section{Results}

We first present briefly some main results
on the {\it minimal model} \cite{selke,holt1,hls,holt2}, illustrating
crucial features by showing Monte Carlo data for typical
equilibrium configurations, see Fig.~2, and for correlation
functions.

\begin{figure}
\begin{center}
\includegraphics[width=0.9\linewidth]{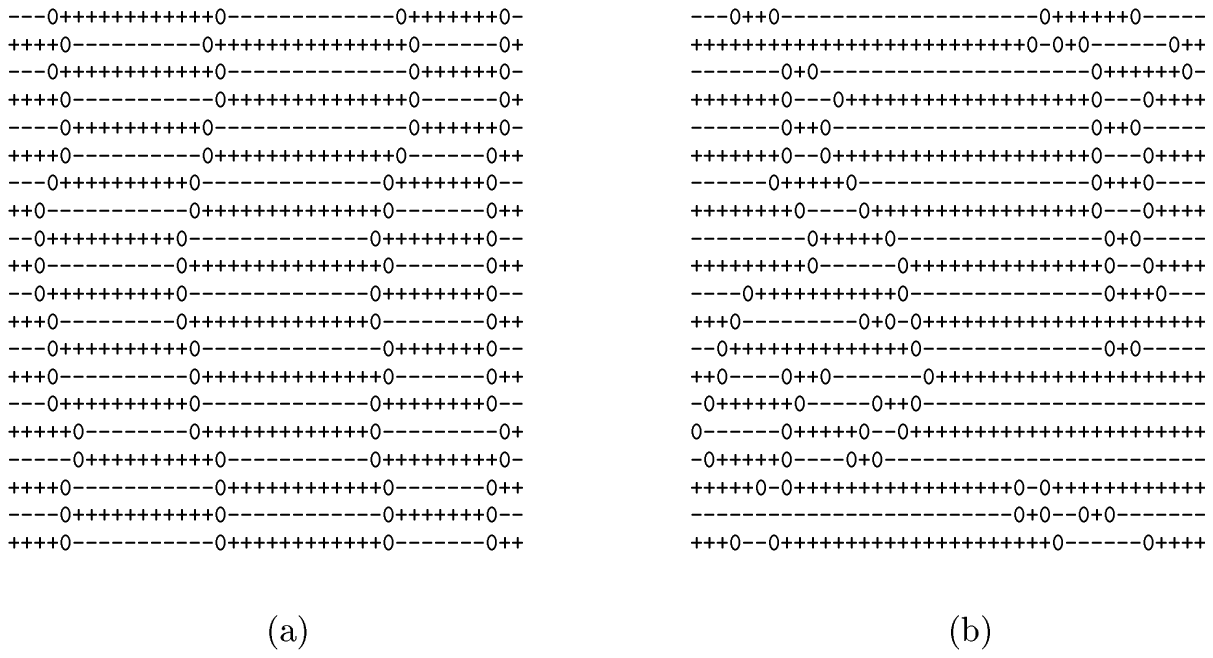}
\end{center}
\caption{Typical Monte Carlo equilibrium configurations of the minimal
model without pinning at $k_BT/|J_a|$= (a) 0.8 and (b) 2.5. Systems
with $L=M=$ 40 were simulated, but only parts are shown. The
transition occurs at $k_BT/|J_a| \approx 1.05$ \cite{selke}.}
\end{figure}

The model with mobile defects, $E_p= 0$, is known to
form, at $T= 0$, straight defect stripes
perpendicular to the chains with arbitrary separation
between the stripes. Correlations along the
chains, $G_1$, oscillate, and the amplitude decays exponentially due to
the large degeneracy of the ground state (compare Fig.~3). On the
other hand, perpendicular to the chains the spins are fully
correlated, $G_2(i,r)= 0.9 (-1)^r$, reflecting the
antiferromagnetic ordering. As temperature $T$ is increased
the stripes will meander, tending to keep, caused by 
entropic repulsion, on average their largest
possible distance. The amplitude of the correlations decays to zero
with distance $r$ algebraically, i.e. there is no long--range
antiferromagnetic order. At a phase transition of first order, the
stripes will break up, accompanied by a pairing of defects.

\begin{figure}
\begin{center}
\begin{minipage}{0.49\linewidth}
\begin{center}
\includegraphics[width=\linewidth]{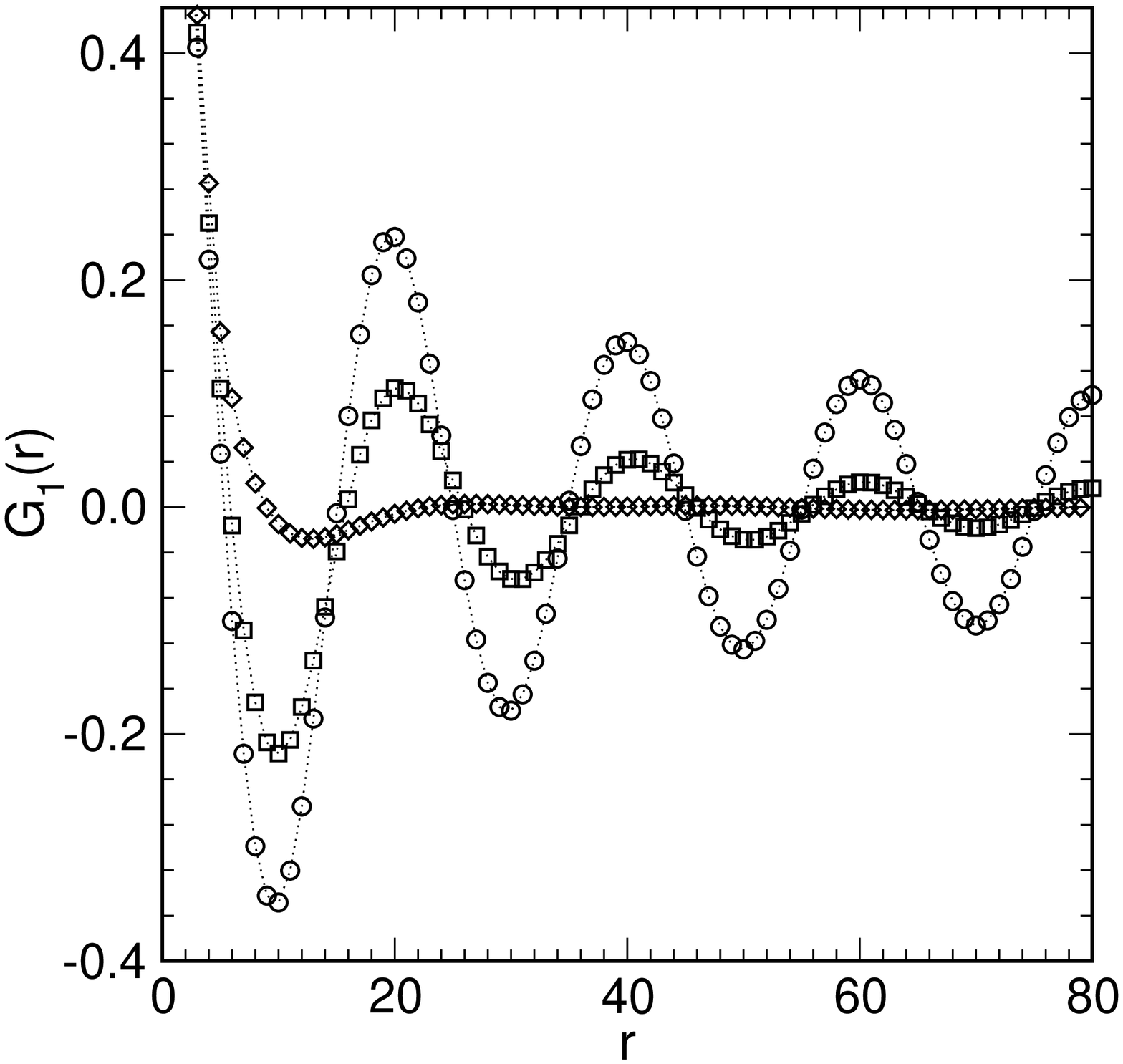}
(a)
\end{center}
\end{minipage}
\begin{minipage}{0.49\linewidth}
\begin{center}
\includegraphics[width=\linewidth]{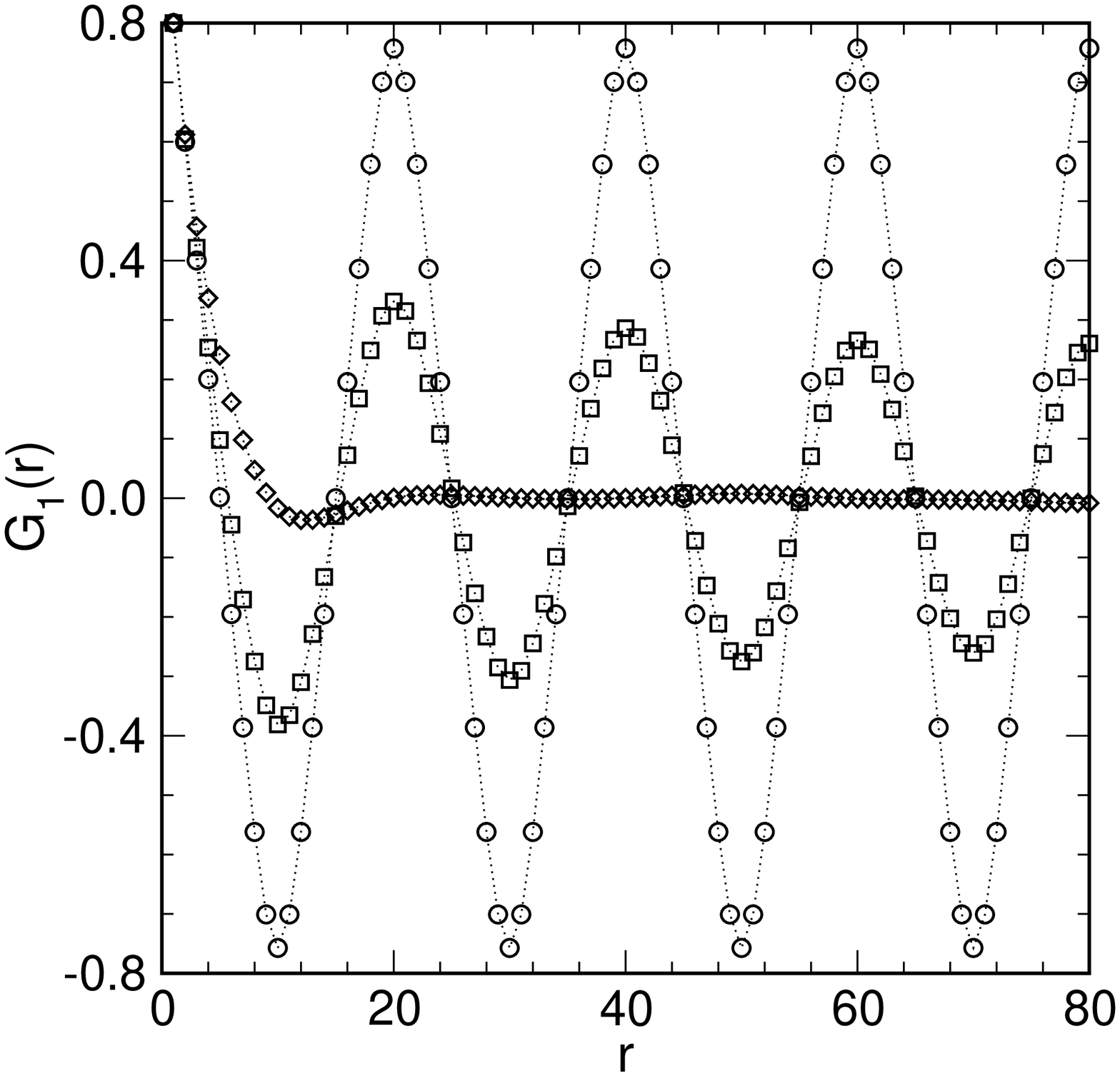}
(b)
\end{center}
\end{minipage}
\end{center}
\caption{Correlation function along the chain direction, $G_1(r)$,
averaging $G_1(i,r)$, Eq.(1), over sites $i$, of the
minimal model (a) without pinning, at $k_BT/|J_a|$= 0.9
(circles), 1.3 (squares), and
1.7 (diamonds), and (b) with periodic pinning lines perpendicular
to the chains at $k_BT/|J_a|$= 1.4 (circles), 1.8 (squares) and
2.0 (diamonds). In the pinned case, the transition occurs at
$k_BT/|J_a| \approx 1.5$. Systems with $L= M= 160$ were simulated.}
\end{figure}

The pairing of the defects, as monitored in the temperature
dependence of the probability of next--nearest
defect pairs $n_d(1)$, results from an attractive effective
interaction between neighbouring defects in a chain. This
interaction, mediated by the magnetic coupling $J_a$, occurs
for strongly fluctuating stripes \cite{selke}.  

Let us now introduce, in the minimal model, a local
pinning, $E_p >0$, of the defects at the sites of straight
equidistant lines perpendicular
to the chains. Then, at low temperatures, the stripes stay close
to these pinning lines and long--range antiferromagnetic order
is observed. The phase transition remains to be of first
order, driven, again, by  the enhanced pairing of
defects \cite{holt1}. Certainly, the transition shifts to
higher temperatures as $E_p$ is increased.

The distinction between the algebraic and long--range
order at low temperatures without and with pinning is
illustrated in Fig.~3. There correlations along the
chains, $G_1$, are depicted at temperatures below, close to, and above
the transition temperature. At low temperatures for $E_p= 0$, the height
of the maxima in the correlations falls off even
at large distance $r$, Fig.~3a, while it reaches quickly
a non--zero constant value in the pinned case $E_p > 0$, see Fig.~3b.

The impact of an external field on the arrangement of the defect
stripes and on the phase transition, both
for mobile and periodically pinned defects, has been discussed
before \cite{hls,holt2}. In particular, at low
temperatures, the defect stripes
are straight at low fields, acquire a zig--zag
structure at larger fields and break up into defect pairs
when further increasing the field.

The influence of random pinning sites on thermal properties
of the minimal model has not been studied in detail yet. Of
course, the limiting case of defects quenched at
random sites is rather trivial, where the amplitude of the
correlations decays exponentially along
and perpendicular to the chain, independent of temperature.
Similarly, periodic quenching leads to full antiferromagnetic
order at any temperature in the minimal model.

\begin{figure}
\begin{center}
\includegraphics[width=0.7\linewidth]{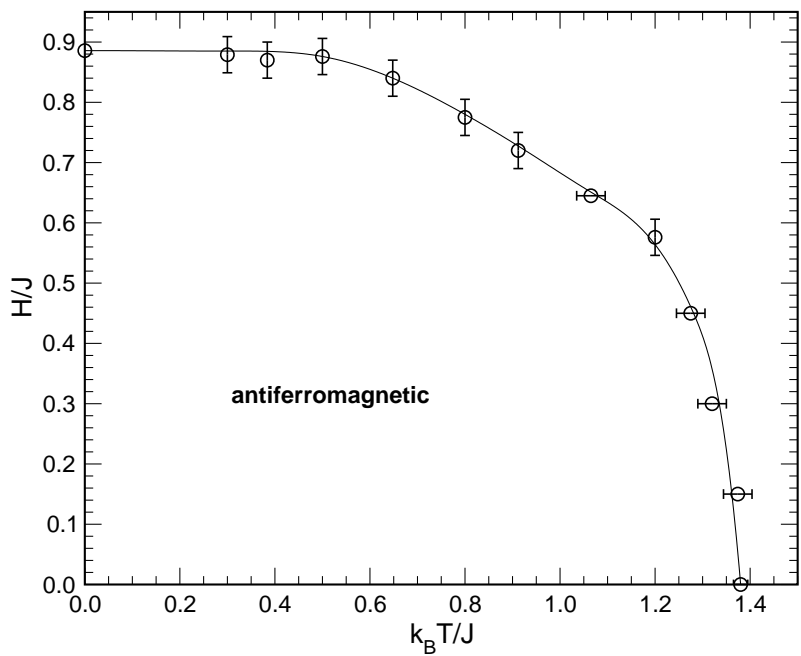}
\end{center}
\caption{Simulated phase diagram of the full model, taking $J_a= -0.3 J$
and $J_0= -6.25J$, with defects quenched
at periodic pinning lines perpendicular
to the chains, $E_p \longrightarrow \infty$. An external
field, $H$, is applied.}
\end{figure}

We now turn to the {\it full model} with $J_a= -0.3 J$ and
$J_0= -6.25 J$. The two spins next to a defect still tend to have
different sign, because $J_0$ is assumed to be rather large. However,
spins between two defects may flip quite easily, with the
flip energy being determined by the intrachain
coupling $J$. Indeed, those quasi one--dimensional spin excitations may mask
the phase transition in some thermodynamic quantities, as may be
seen, for instance, in the specific
heat \cite{selke,holt2}. However, by analysing microscopic quantities
describing the stability of the defect stripes, like the minimal
distance $d_m$ and the probability of encountering defect
pairs separated by merely one spin $n_d(1)$, in the full model
with mobile defects, one observes
again the phase transition driven by the stripe instability
due to defect pairing at a similar temperature as in
the minimal model \cite{selke,holt2}.

Introducing quenched defects in the full
model, quite interesting thermal properties are observed due
to the spin excitations, in contrast to the trivial situation in
the minimal model.

In the case of quenching the defects at periodically
placed lines perpendicular to the chains, we determine, using
standard Monte Carlo techniques, the phase diagram in the
(temperature, field)--plane, as shown in Fig.~4. In
the ground state, when
applying and increasing a field, $H$, the antiferromagnetic
configuration eventually transforms into a (predominantly)
ferromagnetic structure. For sufficiently strong antiferrogmagnetic
coupling at the defects, $J_0$, as it is the case here, one of the two
spins next to a defect will still point against the direction of
the field (so that the antiferromagnetic ordering next to a defect is
still preserved), while
all other spins are aligned parallel to the field. Of
course, further increase of the field, at $T=0$,  will lead to 
full ferromagnetic order even for strong couplings $J_0$ \cite{holt2}.
\begin{figure}
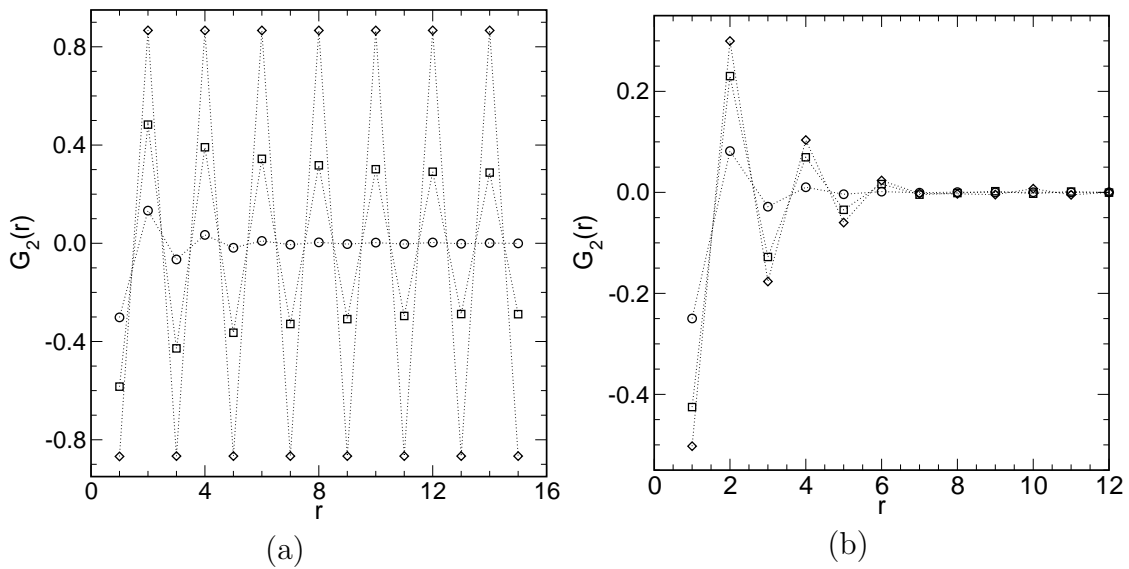

\begin{center}
\begin{minipage}{0.49\linewidth}
\begin{center}
\includegraphics[width=\linewidth]{figure5a}
(a)
\end{center}
\end{minipage}
\begin{minipage}{0.49\linewidth}
\begin{center}
\includegraphics[width=\linewidth]{figure5b}
(b)
\end{center}
\end{minipage}
\end{center}
\caption{Correlation function perpendicular to the chain
direction $G_2(r)$, averaging $G_2(i,r)$, Eq. (2), over sites
$i$, of the full model without external field, $H= 0$, where 
quenched defects are (a) at equidistant
pinning lines perpendicular to the chains, at temperatures
above ($k_BT/J= 1.8$; circles), close to (1.4; squares) and
below (1.0; diamonds)
the transition temperature (see Fig.~4), and (b) at
random sites, above ($k_BT/J= 1.5$; circles), close to (1.0; squares) and
below (0.5; diamonds) the location of
the maximum in the specific heat (see Fig.~6). Lattices with $L=M= 40$ sites
have been simulated.}
\end{figure}

While at zero field, $H= 0$, the phase transition temperature may 
be determined exactly by analytical means \cite {peschel}, Monte
Carlo simulations
are useful to map the entire phase diagram. Here, we obtained
the phase transition line by finite--size analyses on data
for the specific heat as well as the sublattice magnetization. Note
that one encounters, of course, long--range antiferromagnetic ordering in
the low--temperature and low--field phase, as observed, e.g., in the
correlation function perpendicular to the chains, see Fig.~5a.

When quenching defects at random sites, the antiferromagnetic
order seems to be destroyed even at zero temperature.  Ground state
analyses for finite systems indicate that the amplitude
of the correlations, Eqs. (1)
and (2), falls off exponentially with large
distance $r$. However, the behaviour
deviates quantitatively from that
of the minimal model with randomly quenched defects. Indeed, at
$T= 0$, to minimize the energy
of the full model, the antiferromagnetic arrangement of
the spins in adjacent chains, suppressed locally in the
minimal model, may be partly restored by turning over
spins between two consecutive defects.   
\begin{figure}
\begin{center}
\includegraphics[width=0.7\linewidth]{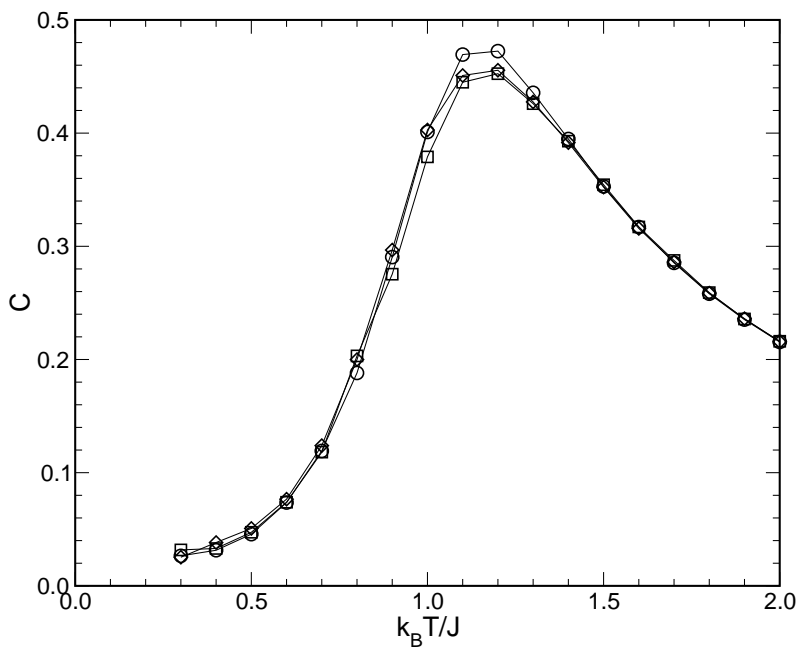}
\end{center}
\caption{Specific heat $C$ vs. temperature for the full model with
defects pinned at random sites. Systems with $L= M$= 20 (circles),
 40 (squares) and 80 (diamonds) have been simulated, averaging over 
ensembles of 40, 20, and 10 realizations.}
\end{figure}
The amplitudes of the simulated spin correlations are observed to
decay rapidly with $r$ even at very low temperatures, as
illustrated in Fig.~5b. In
fact, there is no evidence for 
a transition from an algebraically or even long--range ordered 
antiferromagnetic low--temperature phase to a disordered
phase. Instead, the disordered phase seems to extend down
to $T= 0$.

The absence of a phase transition is also reflected in the
temperature dependence of the specific heat $C$. As depicted in
Fig.~6, $C$ displays a Schottky--type maximum at 
$k_BT/J \approx 1$, whose height depends very weakly on
the system size. Around that temperature, short--range
spin correlations get reduced significantly, due to
the thermally enhanced number of spin flips.

\section{Summary}

In this article, we studied a two--dimensional Ising model
with ferromagnetic interactions between nearest neighbouring
spins along one axis, the chain direction, and somewhat
weaker antiferromagnetic couplings between spins in adjacent
chains. The model has an antiferro- or metamagnetic low--temperature
phase, with a continuous transition to the disordered phase.

Introducing mobile, pinned or quenched defects
with a strong antiferromagnetic interaction between next--nearest
neighbour spins in a chain separated by a defect leads to a variety of
interesting features, both in the minimal and the full models.

In both models, mobile defects tend to form stripes at low
temperatures. At
a phase transition of first order the stripes become
unstable, losing their coherency. The antiferromagnetic ordering
in the low--temperature phase is reduced, characterized by
algebraically decaying spin correlations. 

Long--range order for antiferromagnetic domains may be restored by
pinning the defects at (almost) straight lines perpendicular to
the chains.

Obviously, properties of the minimal model with quenched defects 
do not depend on temperature. By quenching the defects in
the full model at
random sites, the magnetic
ordering at low temperatures is destroyed, and we find no evidence  
for a phase transition.

\vspace{0.5cm}

\section*{Acknowledgements}

We thank B. B\"{u}chner, R. Klingeler, T. Kroll, and
V. L. Pokrovsky for very useful cooperation and information on the topic of
this contribution. Financial support by the Deutsche
Forschungsgemeinschaft under grant No. SE324 is gratefully
acknowledged. One of us (W.S.) thanks Reinhard Folk for numerous
enjoyable discussions on statistical physics and antique books.

\end{document}